\documentclass[useAMS,usenatbib]{mnras}
\usepackage{amsmath,subfig}
\usepackage{graphicx,txfonts,fleqn,enumerate,color,hyperref}
\usepackage{ulem} 

\newcommand*  {\Exp}[1]  {\mathrm{e}^{#1}}

\newcommand*  {\eps}     {\epsilon}

\newcommand   {\sub}[2]  {{#1}_{\mathrm{#2}}}

\newcommand*  {\twovector}[2] {{\begin{pmatrix} $1 \\ $2 \end{pmatrix}}}

\renewcommand {\emph}[1]  {\textit{#1}}

\title[{Discerning Mercury's orbits}]
{{{Stepsize errors in the $N$-body problem: {discerning Mercury's true possible long-term orbits}}}}

\author[David M. Hernandez, Richard E. Zeebe, Sam Hadden]
	{David M. Hernandez$^{1}$\thanks{{{Email: dmhernandez@cfa.harvard.edu} }}, Richard E. Zeebe$^{2}$, Sam Hadden$^{1,3}$ \\ 
	$^1$ Harvard--Smithsonian Center for Astrophysics, 60 Garden St., MS 51, Cambridge, MA 02138, USA \\
	$^2$ School of Ocean and Earth Science and Technology, University of Hawaii at Manoa, 1000 Pope Road, MSB 629, Honolulu, HI 96822, USA\\
	$^3$ Canadian Institute for Theoretical Astrophysics, University of Toronto, 60 St. George Street, Toronto, ON M5S 3H8, Canada\\
	}
\begin{document}

\maketitle

\label{first page}
\begin{abstract} 
{Numerical integrations of the Solar System have been carried out for decades.  Their results have been used, for example, to determine whether the Solar System is chaotic, whether Mercury's orbit is stable, or to help discern Earth's climate history.  We argue that all of the past studies we consider in this work are affected by numerical chaos to different degrees, affecting the possible orbits and instability probability of Mercury, sometimes significantly.  We show how to eliminate the effects of numerical chaos by resolving Mercury's pericentre passage.  We also show that several higher order symplectic maps do not exhibit significant differences in resolving pericentre passage of Mercury (at fixed time step), making their advantages suspect for calculating long-term orbits.  Resolving pericentre passage affects a wide array of orbital numerical studies, like exoplanet studies, studies of the galactic centre, and other $N$-body problems.}
\end{abstract}
\section{Introduction}
\label{sec:first}
$n$-body problems \citep{HH03} describe the motion of point particles interacting through forces that depend on pairwise distances.  In astronomy, they are used to model planetary systems, galaxies, or star clusters, among other problems.  For planetary systems described by conservative Hamiltonian systems, the Lie algebra structure is often used to derive numerical methods known as symplectic methods \citep{hair06}{.  T}he most famous {symplectic method} is arguably the Wisdom--Holman \citep{WH91} method.\footnote{{The Wisdom--Holman map was not derived originally via the Lie algebra framework.}}  This mathematical structure is easily generalized to encompass non-conservative systems \citep[e.g.,][]{Tamayoetal2020}.  Symplectic methods suffer from the well-known and studied roundoff errors and truncation errors.  However, another source of error, arguably less understood, is that
the stepsize chosen is limited by stepsize resonances with the physical
frequencies in the system \citep{WH92}.  The onset of
chaotic behavior can be qualitatively understood in terms of the overlap
of resonances \cite{C79}.  The artificial resonances introduced
by the symplectic map can overlap, introducing an additional source
of unphysical chaos.

\cite{WH92} showed how {stepsize} resonance overlap occurs in a $2$ planet system.  Resonances occur between the numerical time step and the difference of the planets' mean motions.  \cite{RH99} studied numerical chaos in the context of the Stark Problem.  \cite{W15} further showed how resolving pericentre passage avoids effects of stepsize resonances for another two planet system and developed a simple criterion for the timestep required for pericentre resolution.  \cite{Hernandezetal2020} studied statistics of $2$-planet systems, finding accurate statistics when this criterion was respected.

Missing from these works are studies of numerical chaos in more realistic and complex problems like our Solar System, {even if the extension is straightforward}.  Solar System integrations are used to determine the chronology of Earth's climate evolution over geologic timescales \citep[e.g.,][]{montenari18} and the evolution of orbits and chaos.  The effects of resolving pericentre in such studies have not been analyzed, to our knowledge.  Moreover, an analysis of numerical chaos in {higher order and} more sophisticated modern numerical methods beyond those used by \cite{WH92,RH99,W15} is missing.

The aim of this work is to help fill this gap by studying the effects of pericentre resolution in Solar System integrations in the literature.  Additionally, we carry out a study on a wider variety of numerical methods than the works above.  We study three sets of Solar System data, finding they resolve pericentre of Mercury to different degrees.  In particular, we find a data set that does not meet the criterion for pericentre resolution \citep{Abbotetal2021} leads to an evolution of Mercury's eccentricity that is inconsistent at some times with another data set that resolves pericentre to a better degree \citep{Zeebe2015b}.  The paper is organized as follows.  Section \ref{sec:methods} gives an overview of the numerical methods used in this study and carries out a study of how well they resolve pericentre.  Section \ref{sec:experiments} describes Solar System integrations in the literature, and carries out first tests on pericentre resolution in the Solar System.  Section \ref{sec:solarsyst} measures the effects of pericentre resolution on the evolution of Mercury's eccentricity for different data sets.

\section{Methods}
\label{sec:methods}
Consider the $n$-planet problem, which describes perturbed Keplerian motion around a dominant mass.  In this section we describe symplectic \citep{hair06,chann90,H19} and time-reversible \citep{hair06,HB18} maps, or integrators, typically used to study such systems in astrophysics over long time scales.  We use notation described in more detail in \cite{DH17}.  It also more fully gives mathematical background for our work.  A version of the Wisdom--Holman method \citep{WH91,kino91} can be written,
\begin{equation} 
	\Exp{h \sub{\hat{\tilde{H}}}{WH}} = \Exp{\frac{h}{2} \sub{\hat{A}}{}} \Exp{h \sub{\hat{B}}{}}\Exp{\frac{h}{2} \sub{\hat{A}}{}}.
	\label{eq:WH}
\end{equation}
$\hat{A}$ is an operator for function $A$.  $h$ is the fixed timestep.  $\hat{A}$ acts on functions of phase space $f$: $\hat{A} f = \{f,A\}$, where $\{\}$ indicate Poisson brackets.  We have assumed, for simplicity, a conservative system with Hamiltonian $H = A + B$.  The arguments in this paper are easily generalized to nonconservative systems described by first order differential equations, $\dot{x} = (\hat{A} + \hat{B} ) x$ (e.g., \cite{Tamayoetal2020,LR01} ).   $\sub{\tilde{H}}{WH} $ is derived from the BCH formula \citep{hair06}.  $A$ and $B$ can be written in a variety of coordinate systems, and the functions $A$ and $B$ can be chosen differently, so long as $H = A + B$ \citep{HD17}.  $A$ describes Keplerian motion and is integrable.  $B$ includes all other forces and effects in the system.  Eq. \eqref{eq:WH} can{, but need not,} be modified further to produce efficient maps \citep[e.g.,][]{LR01,HD17}.  Another modification is to separate the momentum dependent terms of $B$ (if possible) as $B = B^\prime(q) + S(p)$, where $q$ and $p$ are coordinates and momenta, respectively.  Then, we can write:
\begin{equation} 
	\Exp{h \sub{\hat{\tilde{H}}}{WH}} = \Exp{\frac{h}{2} \sub{\hat{A}}{}} \Exp{\frac{h}{2} \sub{\hat{S}}{}} \Exp{h \sub{\hat{B^\prime}}{}}  \Exp{\frac{h}{2} \sub{\hat{S}}{}}  \Exp{\frac{h}{2} \sub{\hat{A}}{}}.
	\label{eq:WHp}
\end{equation}
Let $\epsilon$ be the ratio of the maximum planetary mass to the dominant mass.  The error of WH is then $\mathcal O(\epsilon h^2)$\footnote{\cite{HD17} use a different convention for $\epsilon$.}.

We denote the next method by WHc, a ``corrected'' \citep{WHT96,W06} version of WH.
\begin{equation} 
	\Exp{h \sub{\hat{\tilde{H}}}{WHc}} =  \Exp{ {\hat{C}}}  \Exp{\frac{h}{2} \sub{\hat{A}}{}} \Exp{h \sub{\hat{B}}{}}\Exp{\frac{h}{2} \sub{\hat{A}}{}}  \Exp{ -{\hat{C}}},
	\label{eq:WHc}
\end{equation}
$\hat{C}$ and $-\hat{C}$ need only be applied at input and output, respectively; this practically incurs no computational penalty.  $\hat{C}$ is constructed to an order $m$, and the error is $\mathcal O(\epsilon h^{m+1} + \epsilon^2 h^2) $.  For $m$ large, the $\mathcal O(\epsilon h^{m+1})$ error can be ignored.  If the function $\delta = 1/24 \{ B, \{B, A\} \}$ is integrable, WHc can be improved further, to WHm.  In particular, $\delta$ is integrable in the common case where $A$ is quadratic in canonical momenta and $B$ depends only on coordinates.  WHm has higher compute cost than WHc and is written, 
\begin{equation} 
	\Exp{h \sub{\hat{\tilde{H}}}{WHm}} = \Exp{ {\hat{C}}} \Exp{\frac{h}{2} \sub{\hat{A}}{}} \Exp{h {\widehat{\sub{B}{} + h^3 \delta}}{}}\Exp{\frac{h}{2} \sub{\hat{A}}{}} \Exp{ -{\hat{C}}}.
	\label{eq:WHM}
\end{equation}
$h^2 \delta < B$ for reasonable timesteps not affected by stepsize chaos.  The error of WHm is $\mathcal O(\epsilon h^{m+1} + \epsilon^2 h^4)$.

Next, we introduce Y, which combines ideas from \cite{Y90} with a WH integrator: 
\begin{equation} 
	\Exp{h \sub{\hat{\tilde{H}}}{Y}} =  \Exp{\alpha h \sub{\hat{\tilde{H}}}{WH}} \Exp{\beta h \sub{\hat{\tilde{H}}}{WH}} \Exp{\alpha h \sub{\hat{\tilde{H}}}{WH}}.
	\label{eq:Y}
\end{equation}
Here, $\alpha = 1/(2 - 2^{1/3})$ and $\beta = -2^{1/3}/(2 - 2^{1/3})$.  The error of Y is $\mathcal O(\epsilon h^4)$, as can be checked numerically.  The integrator package \texttt{HNBody} implements a method equivalent to eqs. \eqref{eq:WHp} and \eqref{eq:Y} in eqs. (8) and (5) in the documentation\footnote{\url{https://www.astro.umd.edu/~hamilton/research/new/RauHam.pdf}}, if certain forces are ignored.

Y can also be corrected, let this method be Yc.  Yc is written,
\begin{equation} 
	\Exp{h \sub{\hat{\tilde{H}}}{Yc}} =  \Exp{ \sub{\hat{C}}{Y}} \Exp{\alpha h \sub{\hat{\tilde{H}}}{WH}} \Exp{\beta h \sub{\hat{\tilde{H}}}{WH}} \Exp{\alpha h \sub{\hat{\tilde{H}}}{WH}} \Exp{ -\sub{\hat{C}}{Y}}.
	\label{eq:Yc}
\end{equation}
The form of $\sub{\hat{C}}{Y}$ has not been published.  The error of Yc is $\mathcal O(\epsilon h^{m+1} + \epsilon^2 h^4)$ (Kevin Rauch, private communication).  For $m$ large, Yc has the same error scaling as WHm.  Note WH, WHc, and WHm (WH methods) share the same operator building blocks, so we might expect similarities in the phase space of orbits computed with them.  Y and Yc (Y methods) {also share similar building blocks}.  When we iterate WH maps, such as \eqref{eq:WH} or \eqref{eq:WHp}, over two or more timesteps, note the coefficient in front of $\hat{A}$ is $h$ (ignoring the first and last operators in the first and last steps, respectively).  For Y methods, the coefficient in front of $\hat{A}$ is $h/2$, so the Kepler problems are solved for slightly shorter times, and are thus better resolved.

Finally, consider LR, referred to as SABA4 by \cite{LR01}:
\begin{equation} 
	\Exp{h \sub{\hat{\tilde{H}}}{LR}} = \Exp{c_1 h \sub{\hat{A}}{}} \Exp{d_1 h \sub{\hat{B}}{}}\Exp{ c_2 h \sub{\hat{A}}{}} \Exp{ d_1 h \sub{\hat{B}}{}} \Exp{ c_1 h \sub{\hat{A}}{}}.
	\label{eq:LR}
\end{equation}
$c_2 \approx 0.577$, $d_1  = 1/2$, and  $c_1 \approx 0.211$.  The error is $\mathcal O(\epsilon h^8 + \eps^2 h^2)$.  The largest coefficient in front of any operator, when we iterate LR, is 0.577, suggesting it has a smaller effective timestep than WH.  Indeed, in experiments with LR, \cite{Hernandezetal2020} found LR can use larger timesteps than WH and still give correct orbital phase space statistics.

We present a solution proposed by \cite{LR01} to use LR with $B$ nonintegrable.  First, assume $B$ can be decomposed into integrable pieces, $B = \sum_{i = 1}^N B_i$.  For $k$ a constant, use the first order accurate substitution, $\Exp{k {\hat{B}_1}{}} \Exp{k {\hat{B}_2}{}}...\Exp{ k {\hat{B}_N}{}}$ , or $\Exp{k {\hat{B}_N}{}} \Exp{k {\hat{B}_{N-1}}{}}...\Exp{ k {\hat{B}_1}{}}$ for $\Exp{ k {\hat{B}}{}}$such that map \eqref{eq:LR} stays symmetric.  These substitutions allow the LR error to remain as $\mathcal O(\epsilon h^8 + \eps^2 h^2)$.  Next, we describe a corrector that reduces the error of LR, at additional expense.  If one of the $B_i$\footnote{It is easy to generalize this argument if more than one $B_i$ satisfy this condition.} depends only on canonical positions, while $A$ is quadratic in the canonical momenta, let $\delta_{\mathrm{LR}} = 1/(2 c) \{ \{A, B_i\}, B_i \}$, where $c \approx 0.003397$ \citep{LR01}.  Then the method, 
\begin{equation} 
	\Exp{h \sub{\hat{\tilde{H}}}{LRm}} = \Exp{-h \sub{\hat{\delta}}{LR}} \Exp{h \sub{\hat{\tilde{H}}}{LR}}  \Exp{h \sub{\hat{\delta}}{LR}},
	\label{eq:LRm}
\end{equation}
eliminates the $\{\{A, B_i\}, B_i \}$ error terms, which scale as $\epsilon^2 h^2$.  {The various methods described here have different merits as far as efficiency goes; i.e., compute cost vs errors committed.  This paper is not preoccupied with the efficiencies; various other works have investigated this \citep[e.g.,][]{Reinetal2019b,Wisdom2018}.

}
\subsection{Pericentre resolution of methods}
\label{sec:nonl}
Nonlinear stability due to pericentre resolution has been investigated for simple problems and methods \citep{RH99,W15}.  Here, we extend this analysis for different WH and Y methods, and investigate pericentre resolution over time.  \cite{W15} proposed that when the stepsize is larger than about $1/16$ of the effective period at pericentre (EPP), pericentre is no longer resolved for several problems.  The EPP is defined as,
\begin{equation}
\tau_f = 2 \pi \sqrt{\frac{(1-e)^3}{1+e} \frac{a^3}{GM} },
\label{eq:peri}
\end{equation}
where $e$ is the eccentricity, $a$ is the semi-major axis, $M$ is the self-gravitating mass, and $G$ is the gravitational constant.

\begin{figure}
	\includegraphics[width=90mm]{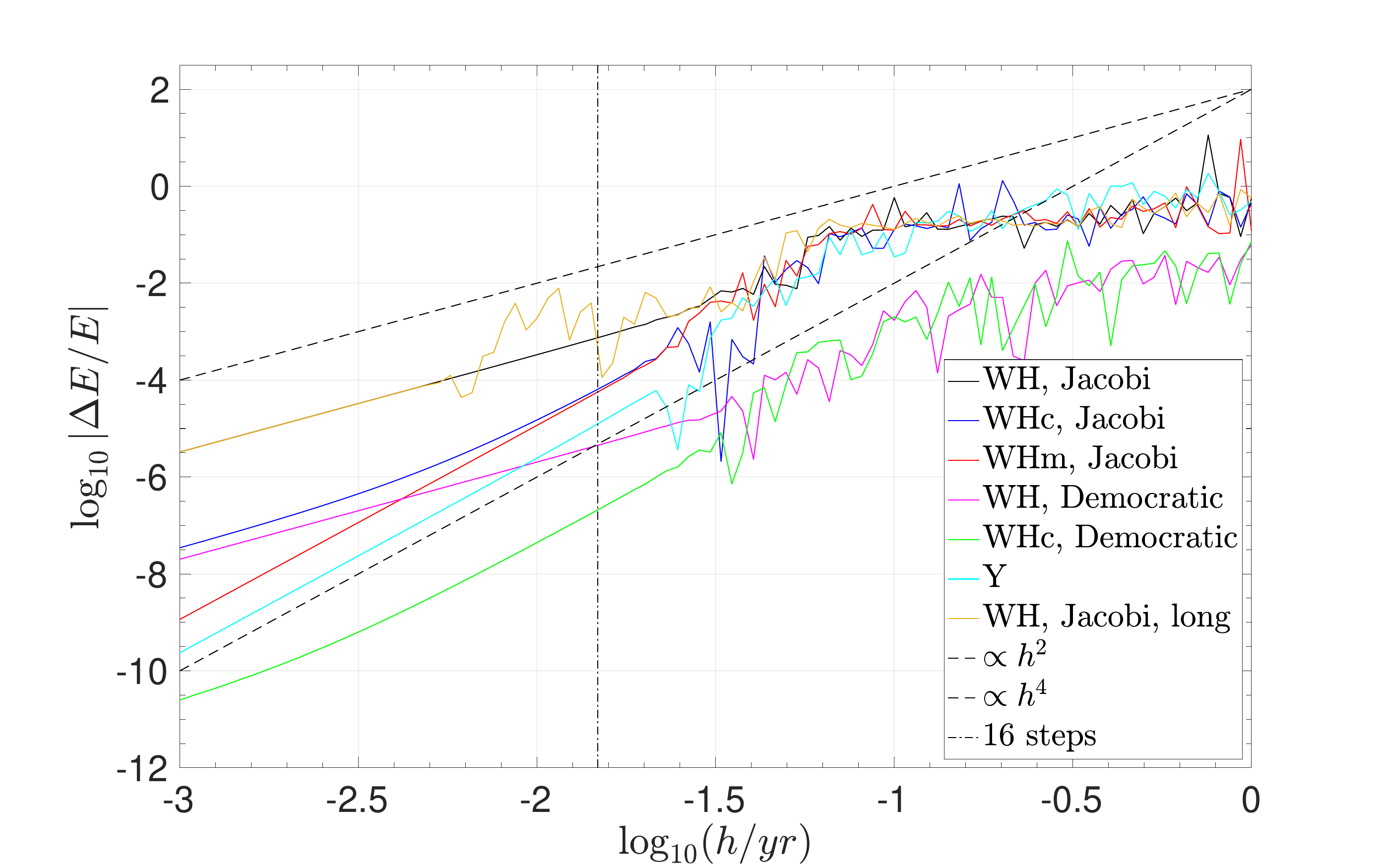}
	\caption{Nonlinear stability of WH and Y integrators.  The two-planet problem is integrated for $1000$ yrs for different stepsizes.  The integrations described by the yellow curve are performed for $10 000$ yrs.  Pericentre resolution for the $1000$ yr integrations is well-described by the vertical line, representing the $\tau_f/16$ criterion.
	\label{fig:stepsize}
  	}
\end{figure}

We consider a first numerical test, the two-planet problem described in \cite{LD00,W15}.  This problem consists of Sun, Jupiter, and Saturn, with masses $(1,~ 1/1047.355,~ 1/3498.5)$ Solar masses.  The inclinations of Jupiter and Saturn are $0$ and $\pi/2$, respectively.  The arguments of perihelion and longitudes of ascending node are set to $0$.  The semi-major axes of Jupiter and Saturn are $5.2$ au and $9.58$ au, respectively.  We use $G = 39.4845$ in units of Solar mass, yr, and au.  This problem is integrated for $1000$ years with various methods for different timesteps.  The implementations of the numerical methods are our own.  The final energy error as a function of timestep is shown in Fig. \ref{fig:stepsize}.
We show different methods, both in Jacobi and Democratic Heliocentric coordinates (for a review, see \cite{HD17}).  {The splitting for Democratic Heliocentric Coordinates and Jacobi coordinates is labelled WHD and WHJ, respectively, in \cite{HD17}.  The ``original form'' gravitating mass for WHJ, as described in \cite{HD17}, is used.  Y is composed of WHJ maps.}    The {orange} curve shows an integration $10$ times longer, for $10 000$ yrs.  For small timesteps, the error of all integrations is dominated by truncation error,  and scales as $h^2$ or $h^4$.  When the error is no longer described well by truncation error, pericentre is no longer resolved.  The criterion $\tau_f/16$ is indicated by a  vertical line.  For the $1000$ yr integrations, this line is a conservative estimate of the transition to the stepsize resonance region.  All methods transition at approximately the same timestep.  This agrees with \cite{Hernandezetal2020}, who compared the probability distribution functions (PDFs) of test particle orbits in the restricted three-body problem.  They found the PDFs broke down at approximately the same time step for a variety of WH methods, where pericentre is no longer resolved.  There are minor differences between the methods, which will be explored in more detail in Section \ref{sec:experiments}.

We see the pericentre resolution requires smaller timesteps for the long integration.  There is no violation of the $\tau_f/16$ rule; what happens in this case is that, for small timesteps, Saturn's eccentricity has significantly increased from $0.950$ at the end of the short simulations, to $0.979$ at the end of the long simulations.  This causes $\tau_f/16 = h$ at $\log_{10} (h/\mathrm{yr}) = -2.4$, which is, to a good approximation, where we see pericentre resolution break down in Fig. \ref{fig:stepsize} for the long integration.
\section{Solar System integrations in the literature}
\label{sec:experiments}
While the simple numerical experiments of Section \ref{sec:nonl} are helpful for getting a grasp on criteria for pericentre resolution, we wish to see the effects of stepsize chaos and diffusion on Solar System integrations, which has not been investigated previously to our knowledge.  We consider three sets of Solar System integrations by different authors which may have different degrees of pericentre resolution of Mercury.

The first set of integrations were carried out by \cite{Zeebe2015b} (Z15).  The Solar System model includes the $8$ planets, Pluto, point mass Newtonian gravity, and general relativistic correctors as described in \cite{ST94}.  The method is Yc, with WH version \eqref{eq:WHp}, as implemented in \texttt{HNBody}.  Jacobi coordinates are used, and $m=6$ for the correctors.  $1600$ Solar System integrations with slightly different initial conditions are performed.  \cite{Zeebe2015b} changes the timestep as Mercury's eccentricity becomes large.  Specifically, for low eccentricities $h = 4$ days is used.  For $0.55 < e < 0.70$, $h = 1$ day is used by default, but see below.  For $0.70 < e < 0.80$, $h = 1/4$ days is used.  For higher eccentricities, an adaptive stepping (non symplectic) method is used.

For eccentricities $(0.2, 0.55, 0.70, 0.80)$, the corresponding $\tau_f/16$ are $(3.59, 1.33, 0.643, 0.35)$ days, respectively.  We have assumed Mercury's semi-major axis remains fixed at $0.387$ au.  To check this assumption, see \cite{Zeebe2015a}, Fig. 1c, where even for high eccentricities nearing $e = 0.8$, Mercury's semi-major axis varies by approximately one part in $10^4$.  It appears the timesteps do not satisfy the $\tau_f/16$ criterion for $e < 0.7$.  However, the criterion is not violated as strongly as we have suggested here: on most of the integrations with high-eccentricity runs, \cite{Zeebe2015b} actually reduces the timestep from $h = 4$ days to $h = 1$ day at eccentricities between $\sim 0.4$ and $\sim 0.5$, rather than $0.55$.  Also, recall the coefficient in front of the $\hat{A}$ operator is half of that of WH methods, which may modify the criterion.  Integrations where the eccentricity of Mercury  becomes larger than $0.55$ are rare; only $1.75\%$ of Z15 integrations have $e > 0.55$.

Next, we consider the integrations by \cite{BR2020}.  $96$ integrations of the Solar System are performed with WHm and $m = 16$ for the corrector.  Their software implementation is called WHCKL \citep{Reinetal2019b}.  They use Jacobi coordinates and $h \approx 8.062$ days.  Their model has the $8$ planets and the general relativistic model of \cite{NR86}, which introduces an error in the mean motion \citep{sah92}.  The extra Hamiltonian terms are only position dependent allowing use of WH version \eqref{eq:WH}.  They state their integrations, ``can be trusted as long as the eccentricity remains moderate, $e\lesssim 0.4$.''  Their stepsize is more than twice that of Z15's, and the difference is even higher for higher Mercury eccentricities.  Thus we expect \cite{BR2020} integrations to have worse pericentre resolution.  \cite{Abbotetal2021} (A+21) perform $1008$ Solar System integrations using the same integration configuration and software package as \cite{BR2020}.  Thus, we can consider A+21 as an extension to the simulations of \cite{BR2020}.

The final set of integrations we consider are those by \cite{LG09} (LG09).  LG09 uses a Solar System model of the $8$ planets, Pluto, general relativity, as suggested by \cite{ST94}, and several other small forces and effects such as the Earth-moon tidal dissipation and the Solar quadrupole moment.  This is the only nonconservative set of integrations.  They carry out $2501$ Solar System integrations using LRm and $h \approx 9.13$ days.  They reduce their timestep for $e > 0.4$, but appear not to state to what value.  Their $B$ is non-integrable, but can be decomposed into integrable parts as discussed in Section \ref{sec:methods}.  Out of the three Solar System data sets we have considered so far, LG09 use the largest timestep.  However, recall from Section \ref{sec:methods} that the effective stepsize for LRm can be considered $\approx 0.577h$, and we conclude that LG09 uses the second largest effective time step for small eccentricities. 

Given their stepsizes, we assume the datasets, ranked from most to least degree of pericentre resolution  are Z15, LG09, and A+21.  We can find some supporting evidence for this already from \cite{Abbotetal2021}, Figure 3.  They compare the Z15, LG09, and A+21 integrations. \cite{Abbotetal2021} compute a Mercury instability probability, finding the datasets, from most to least unstable, are A+21, LG09, and Z15.  Because a failure to resolve pericentre leads to more instability and possibly diffusion, this agrees with our prediction.  They find the instability probability of LG09 and Z15 agree within error bars.  For $t \gtrsim 3.5$ Gyrs, the A+21 instability probabilities are not consistent with the LG09 and Z15 probabilities.  We will revisit this in Section \ref{sec:solarsyst}.

\subsection{Analysis of Z15 integrations}
\label{sec:anal}
To see the effects of resolving Mercury's pericentre on Solar System integrations, we consider the model and integrators used by Z15, described in Section \ref{sec:experiments}.  The initial conditions are taken from the $299$th run, when Mercury's eccentricity has reached $e = 0.84$, so as to minimize the EPP.  The initial conditions are integrated $1000$ yrs using Yc with WH version eq. \eqref{eq:WHp}, and $m = 6$.  Different timesteps are used.  WHc is also used with $m = 6$.  The implementations of \texttt{HNBody} are used.  Mercury's semi-major axis varies by $| \Delta a/a| \le 10^{-4}$ in the integrations.

\begin{figure}
	\includegraphics[width=90mm]{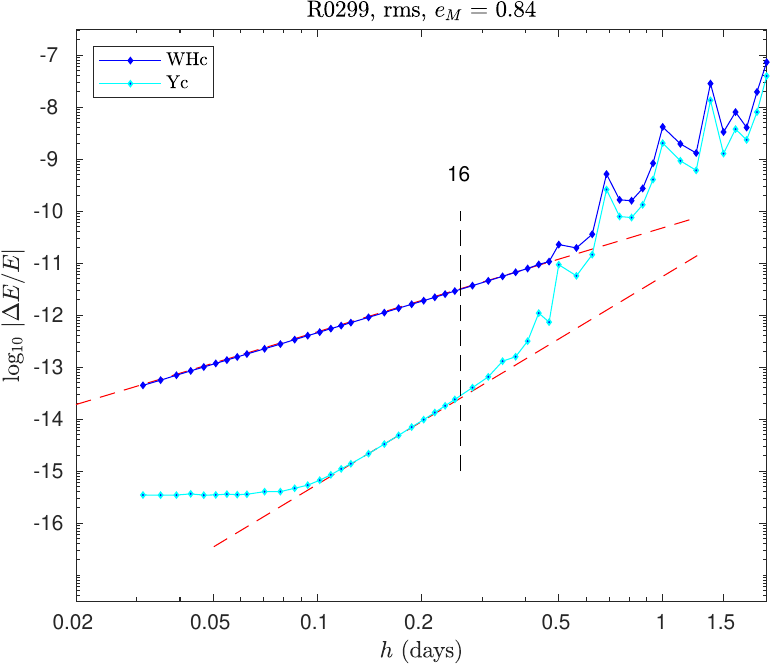}
	\caption{Integration for $1000$ yrs and various timesteps of a Solar System with a highly eccentric Mercury.  Yc and WHc are used, and root-mean-square errors as a function of time are plotted.  A vertical line indicates $\tau_f/16$.  Truncation error dominates over errors from stepsize resonance at small timesteps for WHc.
	\label{fig:dEdt}
  	}
\end{figure}

Root-mean-square energy errors as a function of timestep are plotted in Fig. \ref{fig:dEdt}.  
Also indicated is the $\tau_f/16$ criterion, and slopes $\sim h^2$ and $h^4$.  It appears errors from stepsize resonance present itself at smaller timesteps for Yc, in agreement with comparisons between low and high order methods in \cite{W15}.  In reality, both methods exhibit errors from pericentre passage at approximately the same step, but the WHc errors are too small to be seen on the scale of Fig. \ref{fig:dEdt} for small timesteps.  

\begin{figure}
	\includegraphics[width=90mm]{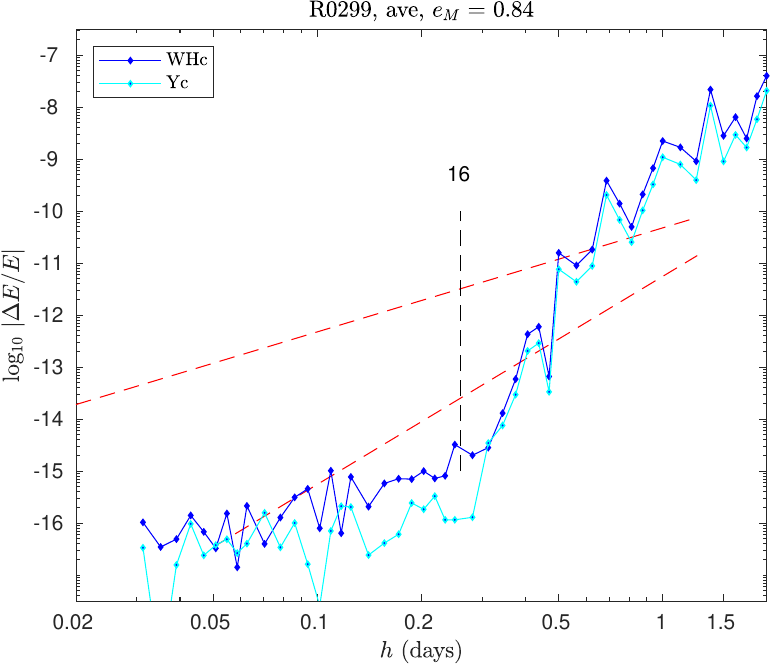}
	\caption{Same as Fig. \ref{fig:dEdt}, but average errors are plotted instead.  WHc and Yc are now seen to transition to the region dominated by roundoff error at similar timesteps predicted by $\tau_f/16$.
	\label{fig:dEdt_ave}
  	}
\end{figure}

To see this, in Fig. \ref{fig:dEdt_ave}, we plot instead the average errors as a function of timestep.
Now, the average errors are dominated by roundoff error for small timesteps, and transition to a region dominated by stepsize resonances at approximately the same time step, in agreement with the $\tau_f/16$ criterion.  We have also run this problem without Mercury and verified that the effects of stepsize resonances disappear for the timestep ranges and the scale in error of Fig. \ref{fig:dEdt_ave}.

In our experiments, the evolution of stepsize chaos with time is small and insignificant.  This is consistent with \cite{Hernandezetal2020}, Fig. 10.  They studied the variance of the Jacobi constant error over time for restricted three-body problems.  This variance grew linearly for symplectic integrators with moderate timesteps.  For small timesteps, there was no clear trend in the evolution of the variance.
\section{Influence of pericentre resolution on Mercury's eccentricity}
\label{sec:solarsyst}

\begin{figure}
\includegraphics[width=90mm]{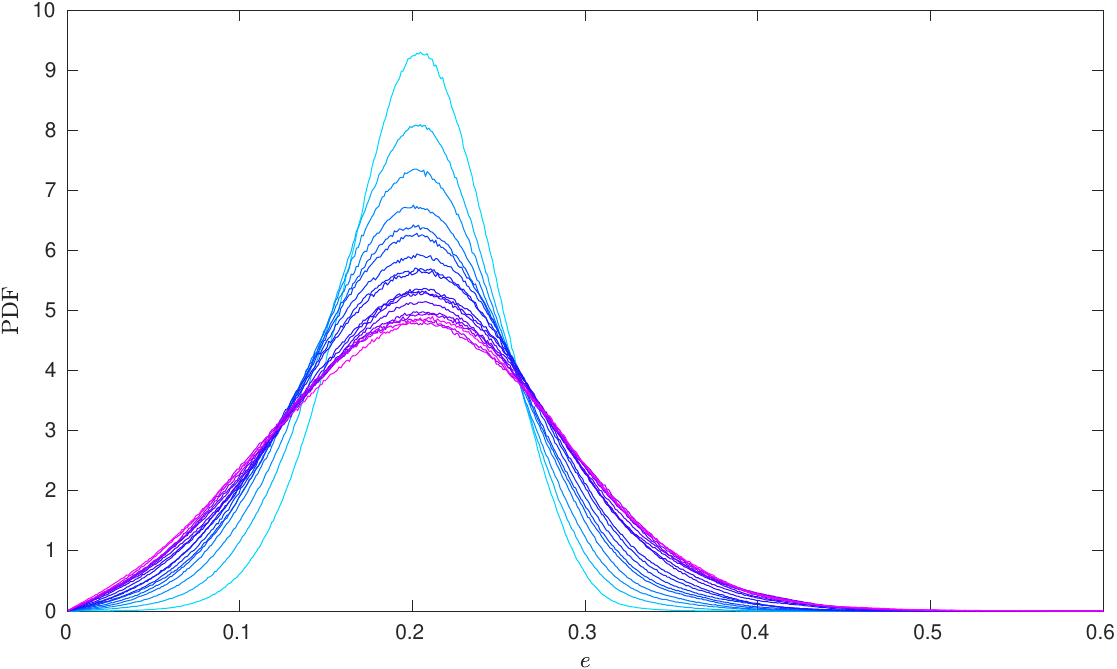}
	\caption{PDFs of Mercury's eccentricity over time from the Z15 data.  A PDF is plotted every $250$ Myrs.  The first $250$ Myr time slice is not shown. {The colors towards the purple end of the spectrum indicate longer integration times and show Mercury's eccentricity diffusing.}
	\label{fig:mercurypdf}
  	}
\end{figure}

The diffusion of Mercury's eccentricity is one mechanism for encountering Solar System instabilities \citep{Laskar08}.  Here, we compare ensembles of $N$-body integrations with different time steps to explore how resolution of Mercury's EPP influences this chaotic diffusion; specifically, we will compare A+21 and Z15.  In Fig. \ref{fig:mercurypdf}, we compute the probability density function (PDF) of Mercury's eccentricity every $250$ Myrs, using the Z15 data sets.  The PDF data for the first $250$ Myrs are discarded.  The colors towards the purple end of the spectrum indicate longer integration times and show Mercury's eccentricity diffusing \citep{BMH2015}.  A sampling rate of $\sim 25$ kyrs is used and the eccentricities are sorted into $700$ bins between eccentricities $0$ to $0.7$.  This figure can be compared to \cite{Laskar08}, Figure 6.  

To quantify the time evolution of Mercury's eccentricity in the two ensembles of integrations, we fit a two-parameter Rice distribution, with a probability density given by
\begin{equation}
f_{\eta,m} (x) = \frac{x}{\eta^2} \exp \left( - \frac{x^2 + m^2}{2 \eta^2} \right) I_0 \left( \frac{x m}{\eta^2} \right),
\label{eq:ricedist}
\end{equation}
where $I_0$ is the modified Bessel function of the first kind, to the distribution of Mercury's eccentricity at different times in each ensemble.  Note we are reusing the variable $m$, so that our notation is consistent with \cite{Laskar08}.

 We derive maximum-likelihood estimates for the distribution parameters, $m$  and $\eta^2$, by maximizing the likelihood function ${\cal L}(m,\eta^2|\{e_i\}_{t=T}) \propto \prod_{i=1}^N f_{\eta,m}(e_i)$ 
 for the set of $N$ values of Mercury's eccentricity, $\{e_i\}_{t=T}$, at a given time, $T$, among the simulations in an ensemble.    Note the samples obtained in this way are composed of approximately $1008$ and $1600$ eccentricity values per time slice for A+21 and Z15, respectively (simulations that experienced instabilities prior to the time $T$ are removed from the ensembles before fitting for distribution parameters).  A Z15 data point at $\approx 2.6$ Gyrs is missing as output was not saved at that time.
 
We also obtain estimate uncertainties in the inferred parameters by computing the Hessian of the log-likelihood function at the maximum likelihood value of the parameters and using the Laplace approximation. In other words, we assume our maximum likelihood estimates of the parameters $m$ and $\eta^2$ are approximately Gaussian-distributed about the true values with  covariance matrix $\Sigma =-[\nabla^2_{(m,\eta^2) }\ln\mathcal{L}]^{-1}$.  
 
 \begin{figure}
	\includegraphics[width=90mm]{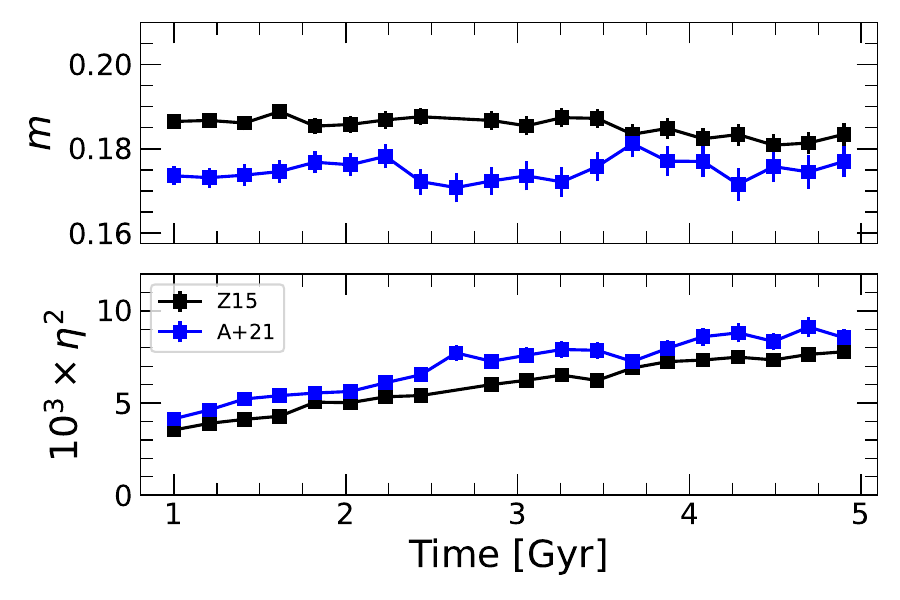}
	\caption{Comparison of the Z15 (black) and A+21 (blue) PDFs of Mercury's eccentricity as a function of time.
	Each panel shows a parameter of the Rice distribution, Equation \eqref{eq:ricedist}, fitted to the sample of Mercury's eccentricity values taken from each ensemble of simulations at the plotted times.
	Error bars indicate estimated $1\sigma$ uncertainties in parameter values.
	\label{fig:RiceParm}
  	}
\end{figure}

 \begin{figure}
	\includegraphics[width=90mm]{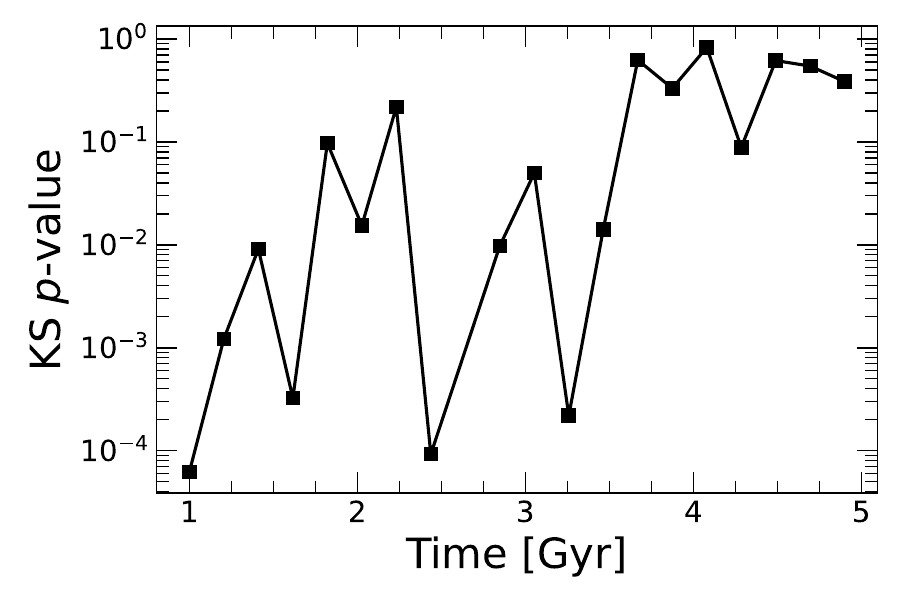}
	\caption{
	$p$-values from a two-sample KS test comparing the A+21 and Z15 simulation PDFs.
	\label{fig:ks-pvalue}
  	}
\end{figure}
The parameter $\eta$ is a proxy for the Rician width.  $\eta^2$ is statistically significantly larger for A+21, as a function of time.  This may be consistent with A+21, who find in their Figure 3 that A+21 has higher instability probabilities than Z15 as a function of time.  They demonstrate their instability probabilities are correlated with eccentricity range.  $\eta$ and eccentricity range are likely correlated as well.  $m$, which indicates the centering of the PDFs, is significantly smaller for A+21 for almost all time.

A two-sample KS test between the A+21 and Z15 data provides a more direct test of the compatibility of Mercury's eccentricity distribution in the two simulation ensembles without imposing any assumptions about an underlying functional form.  In Fig. \ref{fig:ks-pvalue}, we plot $p$-values from two-sample KS tests as a function of time.  The $p$-values are small {($<0.05$)} for most data at time $t < 3.5$ Gyrs, indicating inconsistent simulations.  However, for time $t > 3.5$ Gyrs, the $p$-values are large {($>0.05$)}, indicating consistent PDFs between A+21 and Z15.  {One might wonder how sensitive our KS $p$-values are to our choice of time slices.  To explore this, we have recalculated Fig. \ref{fig:ks-pvalue} twice, first shifting the times by $-100$ Myr and then by $+1$ Myr.  There are variations in the $p$-values, but the general trends remain: at times $> 3.5  $ Gyr, the PDFs are mostly consistent, while at earlier times they are mostly inconsistent}.  This is the opposite trend observed by A+21 in their Figure 3, who find the instability probabilities of A+21 and Z15 become inconsistent for $t \gtrsim 3.5$ Gyrs.  

\section{Discussion}
\label{sec:conc}
We studied the effects of numerical chaos on Solar System integrations in the literature.  We find significant differences in the evolution of Mercury's eccentricity between the data sets of \cite{Zeebe2015b} and \cite{Abbotetal2021}, and we postulate this is due to different levels of pericentre resolution.  Our results are also supported by the study of \cite{Abbotetal2021}, which compares Solar System instability probabilities between different data sets.  We conclude that a failure to resolve pericentre can lead to unphysical orbits in dynamical systems in astronomy.  However, we believe the comparison between the A+21 and Z15 data sets can be strengthened by increasing the number of integrations in each data set.  Stepsize resonance errors can be avoided by adequately resolving the effective period at pericentre, Eq. \eqref{eq:peri}.  For a wide range of numerical methods (except a \cite{LR01} method), the tests here and in other works \citep{W15,Hernandezetal2020} suggest a time step of $\tau_f/16$ adequately avoids stepsize chaos, although this criterion may need slight modification for some problems \citep{W15}.

We note none of the Solar System data sets we have considered here strictly resolve pericentre; {some are closer to resolving it than others, which explains the different instability probabilities of \cite{Abbotetal2021}.}  It may be that Mercury's eccentricity evolution has not converged yet and looks different for an integration that uses smaller stepsizes than any considered here.  Such stepsizes might better avoid numerical chaos and diffusion.  Also note that the energy of a Keplerian orbit is independent of the eccentricity, so energy errors do not directly tell us about the eccentricity evolution of Mercury.

Lower order symplectic integrators can resolve Mercury's pericentre at lower computational cost compared to several higher order methods.  This is a disadvantage of using higher order methods instead of lower order methods for Solar System calculations, and raises more questions about the usefulness of higher order symplectic integrators for planetary dynamics \citep{Hetal2021}.

Finally, we mention that pericentre resolution is not a problem unique to planetary systems.  Indeed, we expect resolving pericentre to be important in a variety of $N$-body systems, such as stellar clusters and galaxies.  Future work can explore this issue.

\section{Acknowledgements}
We thank Matt Holman and Jack Wisdom for {suggestions.}

\section{Data Availability}
The data underlying this article will be shared on reasonable request to the corresponding author.

\bibliographystyle{mnras}
\bibliography{paper}

\begin{thebibliography}{}
\makeatletter
\relax
\def\mn@urlcharsother{\let\do\@makeother \do\$\do\&\do\#\do\^\do\_\do\%\do\~}
\def\mn@doi{\begingroup\mn@urlcharsother \@ifnextchar [ {\mn@doi@}
  {\mn@doi@[]}}
\def\mn@doi@[#1]#2{\def\@tempa{#1}\ifx\@tempa\@empty \href
  {http://dx.doi.org/#2} {doi:#2}\else \href {http://dx.doi.org/#2} {#1}\fi
  \endgroup}
\def\mn@eprint#1#2{\mn@eprint@#1:#2::\@nil}
\def\mn@eprint@arXiv#1{\href {http://arxiv.org/abs/#1} {{\tt arXiv:#1}}}
\def\mn@eprint@dblp#1{\href {http://dblp.uni-trier.de/rec/bibtex/#1.xml}
  {dblp:#1}}
\def\mn@eprint@#1:#2:#3:#4\@nil{\def\@tempa {#1}\def\@tempb {#2}\def\@tempc
  {#3}\ifx \@tempc \@empty \let \@tempc \@tempb \let \@tempb \@tempa \fi \ifx
  \@tempb \@empty \def\@tempb {arXiv}\fi \@ifundefined
  {mn@eprint@\@tempb}{\@tempb:\@tempc}{\expandafter \expandafter \csname
  mn@eprint@\@tempb\endcsname \expandafter{\@tempc}}}

\bibitem[\protect\citeauthoryear{{Abbot}, {Webber}, {Hadden}  \&
  {Weare}}{{Abbot} et~al.}{2021}]{Abbotetal2021}
{Abbot} D.~S.,  {Webber} R.~J.,  {Hadden} S.,   {Weare} J.,  2021, arXiv
  e-prints, \href {https://ui.adsabs.harvard.edu/abs/2021arXiv210609091A} {p.
  arXiv:2106.09091}

\bibitem[\protect\citeauthoryear{{Batygin}, {Morbidelli}  \&
  {Holman}}{{Batygin} et~al.}{2015}]{BMH2015}
{Batygin} K.,  {Morbidelli} A.,   {Holman} M.~J.,  2015, \mn@doi [ApJ]
  {10.1088/0004-637X/799/2/120}, \href
  {https://ui.adsabs.harvard.edu/abs/2015ApJ...799..120B} {799, 120}

\bibitem[\protect\citeauthoryear{{Brown} \& {Rein}}{{Brown} \&
  {Rein}}{2020}]{BR2020}
{Brown} G.,  {Rein} H.,  2020, \mn@doi [Research Notes of the American
  Astronomical Society] {10.3847/2515-5172/abd103}, \href
  {https://ui.adsabs.harvard.edu/abs/2020RNAAS...4..221B} {4, 221}

\bibitem[\protect\citeauthoryear{{Channell} \& {Scovel}}{{Channell} \&
  {Scovel}}{1990}]{chann90}
{Channell} P.~J.,  {Scovel} C.,  1990, \mn@doi [Nonlinearity]
  {10.1088/0951-7715/3/2/001}, 3, 231

\bibitem[\protect\citeauthoryear{{Chirikov}}{{Chirikov}}{1979}]{C79}
{Chirikov} B.~V.,  1979, \mn@doi [Physics Reports]
  {10.1016/0370-1573(79)90023-1}, \href
  {http://adsabs.harvard.edu/abs/1979PhR....52..263C} {52, 263}

\bibitem[\protect\citeauthoryear{{Dehnen} \& {Hernandez}}{{Dehnen} \&
  {Hernandez}}{2017}]{DH17}
{Dehnen} W.,  {Hernandez} D.~M.,  2017, \mn@doi [\mnras]
  {10.1093/mnras/stw2758}, \href
  {http://adsabs.harvard.edu/abs/2017MNRAS.465.1201D} {465, 1201}

\bibitem[\protect\citeauthoryear{{Hairer}, {Lubich}  \& {Wanner}}{{Hairer}
  et~al.}{2006}]{hair06}
{Hairer} E.,  {Lubich} C.,   {Wanner} G.,  2006, {Geometrical Numerical
  Integration}, 2nd edn.
Springer Verlag, Berlin

\bibitem[\protect\citeauthoryear{{Heggie} \& {Hut}}{{Heggie} \&
  {Hut}}{2003}]{HH03}
{Heggie} D.,  {Hut} P.,  2003, {The Gravitational Million-Body Problem: A
  Multidisciplinary Approach to Star Cluster Dynamics}.
Cambridge University Press

\bibitem[\protect\citeauthoryear{{Hernandez}}{{Hernandez}}{2019}]{H19}
{Hernandez} D.~M.,  2019, \mn@doi [\mnras] {10.1093/mnras/stz884}, \href
  {https://ui.adsabs.harvard.edu/abs/2019MNRAS.486.5231H} {486, 5231}

\bibitem[\protect\citeauthoryear{{Hernandez} \& {Bertschinger}}{{Hernandez} \&
  {Bertschinger}}{2018}]{HB18}
{Hernandez} D.~M.,  {Bertschinger} E.,  2018, \mn@doi [MNRAS]
  {10.1093/mnras/sty184}, \href
  {http://adsabs.harvard.edu/abs/2018MNRAS.475.5570H} {475, 5570}

\bibitem[\protect\citeauthoryear{{Hernandez} \& {Dehnen}}{{Hernandez} \&
  {Dehnen}}{2017}]{HD17}
{Hernandez} D.~M.,  {Dehnen} W.,  2017, \mn@doi [\mnras]
  {10.1093/mnras/stx547}, \href
  {http://adsabs.harvard.edu/abs/2017MNRAS.468.2614H} {468, 2614}

\bibitem[\protect\citeauthoryear{{Hernandez}, {Hadden}  \&
  {Makino}}{{Hernandez} et~al.}{2020}]{Hernandezetal2020}
{Hernandez} D.~M.,  {Hadden} S.,   {Makino} J.,  2020, \mn@doi [\mnras]
  {10.1093/mnras/staa388}, \href
  {https://ui.adsabs.harvard.edu/abs/2020MNRAS.493.1913H} {493, 1913}

\bibitem[\protect\citeauthoryear{{Hernandez}, {Agol}, {Holman}  \&
  {Hadden}}{{Hernandez} et~al.}{2021}]{Hetal2021}
{Hernandez} D.~M.,  {Agol} E.,  {Holman} M.~J.,   {Hadden} S.,  2021, \mn@doi
  [Research Notes of the American Astronomical Society]
  {10.3847/2515-5172/abf4e3}, \href
  {https://ui.adsabs.harvard.edu/abs/2021RNAAS...5...77H} {5, 77}

\bibitem[\protect\citeauthoryear{{Kinoshita}, {Yoshida}  \&
  {Nakai}}{{Kinoshita} et~al.}{1991}]{kino91}
{Kinoshita} H.,  {Yoshida} H.,   {Nakai} H.,  1991, Celest. Mech. Dyn. Astron.,
  50, 59

\bibitem[\protect\citeauthoryear{{Laskar}}{{Laskar}}{2008}]{Laskar08}
{Laskar} J.,  2008, \mn@doi [\icarus] {10.1016/j.icarus.2008.02.017}, \href
  {https://ui.adsabs.harvard.edu/abs/2008Icar..196....1L} {196, 1}

\bibitem[\protect\citeauthoryear{{Laskar} \& {Gastineau}}{{Laskar} \&
  {Gastineau}}{2009}]{LG09}
{Laskar} J.,  {Gastineau} M.,  2009, \mn@doi [\nat] {10.1038/nature08096},
  \href {https://ui.adsabs.harvard.edu/abs/2009Natur.459..817L} {459, 817}

\bibitem[\protect\citeauthoryear{{Laskar} \& {Robutel}}{{Laskar} \&
  {Robutel}}{2001}]{LR01}
{Laskar} J.,  {Robutel} P.,  2001, Celestial Mechanics and Dynamical Astronomy,
  \href {https://ui.adsabs.harvard.edu/abs/2001CeMDA..80...39L} {80, 39}

\bibitem[\protect\citeauthoryear{{Levison} \& {Duncan}}{{Levison} \&
  {Duncan}}{2000}]{LD00}
{Levison} H.~F.,  {Duncan} M.~J.,  2000, \mn@doi [AJ] {10.1086/301553}, \href
  {http://adsabs.harvard.edu/abs/2000AJ....120.2117L} {120, 2117}

\bibitem[\protect\citeauthoryear{Montenari}{Montenari}{2018}]{montenari18}
Montenari M.,  2018, {(Editor) Stratigraphy \& Timescales: Cyclostratigraphy
  and Astrochronology in 2018}.
~ Vol. 3, Elsevier

\bibitem[\protect\citeauthoryear{{Nobili} \& {Roxburgh}}{{Nobili} \&
  {Roxburgh}}{1986}]{NR86}
{Nobili} A.,  {Roxburgh} I.~W.,  1986, in {Kovalevsky} J.,  {Brumberg} V.~A.,
  eds,  Vol. 114, Relativity in Celestial Mechanics and Astrometry. High
  Precision Dynamical Theories and Observational Verifications. p.~105

\bibitem[\protect\citeauthoryear{{Rauch} \& {Holman}}{{Rauch} \&
  {Holman}}{1999}]{RH99}
{Rauch} K.~P.,  {Holman} M.,  1999, \mn@doi [\aj] {10.1086/300720}, \href
  {https://ui.adsabs.harvard.edu/abs/1999AJ....117.1087R} {117, 1087}

\bibitem[\protect\citeauthoryear{{Rein}, {Tamayo}  \& {Brown}}{{Rein}
  et~al.}{2019}]{Reinetal2019b}
{Rein} H.,  {Tamayo} D.,   {Brown} G.,  2019, \mn@doi [\mnras]
  {10.1093/mnras/stz2503}, \href
  {https://ui.adsabs.harvard.edu/abs/2019MNRAS.489.4632R} {489, 4632}

\bibitem[\protect\citeauthoryear{{Saha} \& {Tremaine}}{{Saha} \&
  {Tremaine}}{1992}]{sah92}
{Saha} P.,  {Tremaine} S.,  1992, \mn@doi [AJ] {10.1086/116347}, 104, 1633

\bibitem[\protect\citeauthoryear{{Saha} \& {Tremaine}}{{Saha} \&
  {Tremaine}}{1994}]{ST94}
{Saha} P.,  {Tremaine} S.,  1994, \mn@doi [\aj] {10.1086/117210}, \href
  {https://ui.adsabs.harvard.edu/abs/1994AJ....108.1962S} {108, 1962}

\bibitem[\protect\citeauthoryear{{Tamayo}, {Rein}, {Shi}  \&
  {Hernandez}}{{Tamayo} et~al.}{2020}]{Tamayoetal2020}
{Tamayo} D.,  {Rein} H.,  {Shi} P.,   {Hernandez} D.~M.,  2020, \mn@doi
  [\mnras] {10.1093/mnras/stz2870}, \href
  {https://ui.adsabs.harvard.edu/abs/2020MNRAS.491.2885T} {491, 2885}

\bibitem[\protect\citeauthoryear{{Wisdom}}{{Wisdom}}{2006}]{W06}
{Wisdom} J.,  2006, \mn@doi [AJ] {10.1086/500829}, \href
  {http://adsabs.harvard.edu/abs/2006AJ....131.2294W} {131, 2294}

\bibitem[\protect\citeauthoryear{{Wisdom}}{{Wisdom}}{2015}]{W15}
{Wisdom} J.,  2015, \mn@doi [AJ] {10.1088/0004-6256/150/4/127}, \href
  {http://adsabs.harvard.edu/abs/2015AJ....150..127W} {150, 127}

\bibitem[\protect\citeauthoryear{{Wisdom}}{{Wisdom}}{2018}]{Wisdom2018}
{Wisdom} J.,  2018, \mn@doi [\mnras] {10.1093/mnras/stx2906}, \href
  {https://ui.adsabs.harvard.edu/abs/2018MNRAS.474.3273W} {474, 3273}

\bibitem[\protect\citeauthoryear{{Wisdom} \& {Holman}}{{Wisdom} \&
  {Holman}}{1991}]{WH91}
{Wisdom} J.,  {Holman} M.,  1991, \mn@doi [AJ] {10.1086/115978}, 102, 1528

\bibitem[\protect\citeauthoryear{{Wisdom} \& {Holman}}{{Wisdom} \&
  {Holman}}{1992}]{WH92}
{Wisdom} J.,  {Holman} M.,  1992, \mn@doi [\aj] {10.1086/116378}, \href
  {https://ui.adsabs.harvard.edu/abs/1992AJ....104.2022W} {104, 2022}

\bibitem[\protect\citeauthoryear{{Wisdom}, {Holman}  \& {Touma}}{{Wisdom}
  et~al.}{1996}]{WHT96}
{Wisdom} J.,  {Holman} M.,   {Touma} J.,  1996, Fields Institute
  Communications, Vol.~10, p.~217, \href
  {http://adsabs.harvard.edu/abs/1996FIC....10..217W} {10, 217}

\bibitem[\protect\citeauthoryear{{Yoshida}}{{Yoshida}}{1990}]{Y90}
{Yoshida} H.,  1990, Physics Letters A, 150, 262

\bibitem[\protect\citeauthoryear{{Zeebe}}{{Zeebe}}{2015a}]{Zeebe2015a}
{Zeebe} R.~E.,  2015a, \mn@doi [\apj] {10.1088/0004-637X/798/1/8}, \href
  {https://ui.adsabs.harvard.edu/abs/2015ApJ...798....8Z} {798, 8}

\bibitem[\protect\citeauthoryear{{Zeebe}}{{Zeebe}}{2015b}]{Zeebe2015b}
{Zeebe} R.~E.,  2015b, \mn@doi [ApJ] {10.1088/0004-637X/811/1/9}, \href
  {https://ui.adsabs.harvard.edu/abs/2015ApJ...811....9Z} {811, 9}

\makeatother
\end{thebibliography}

\end{document}